\documentclass[twocolumn]{article}
\usepackage[margin=2cm]{geometry}
\usepackage{graphicx}
\usepackage{amsmath, amssymb}
\usepackage{hyperref}
\usepackage{cite}
\usepackage{caption}
\usepackage{subcaption}
\usepackage{bm}
\usepackage{algorithm}
\usepackage{algpseudocode}
\usepackage{listings}
\usepackage{color}
\usepackage{authblk}
\usepackage{orcidlink}
\title{Mitigating Frequency Learning Bias in Quantum Models via Multi‑Stage Residual Learning}

\author{Ammar Daskin \orcidlink{0000-0002-1497-5031}}
\affil{
Department of Computer Engineering\\
Istanbul Medeniyet University,
Istanbul, Turkiye, 34000\\
email: adaskin25@gmail.com
}
\date{}

\begin{document}

\maketitle

\begin{abstract}
Quantum machine learning models based on parameterized circuits can be viewed as Fourier series approximators. However, they often struggle to learn functions with multiple frequency components, particularly high‑frequency or non-dominant ones - a phenomenon we term the \textit{quantum Fourier parameterization bias}. Inspired by recent advances in classical Fourier neural operators (FNOs), we adapt the multi-stage residual learning idea to the quantum domain, iteratively training additional quantum modules on the residuals of previous stages.  We evaluate our method on a synthetic benchmark composed of spatially localized frequency components with diverse envelope shapes (Gaussian, Lorentzian, triangular). Systematic experiments show that the number of qubits, the encoding scheme, and residual learning are all crucial for resolving multiple frequencies; residual learning alone can improve test MSE significantly over a single‑stage baseline trained for the same total number of epochs. Our work provides a practical framework for enhancing the spectral expressivity of quantum models and offers new insights into their frequency-learning behavior.
\end{abstract}

% ----- INTRODUCTION -----
\section{Introduction}
\label{sec:intro}

\subsection{Motivation}
\paragraph{Varying Temporal and Spatial Frequency Contents in Data Analysis.}
Many scientific and engineering tasks rely on understanding how data vary in time or space. This variation is captured by its frequency content: low frequencies represent broad, slowly changing trends, while high frequencies encode rapid fluctuations or fine details. In time‑series analysis, for example, low‑frequency components reflect long‑term patterns, and high‑frequency components capture noise or short‑term events (see e.g. the book \cite{pachori2023time}). In images, low spatial frequencies correspond to coarse structures, and high spatial frequencies to edges and textures ( e.g. see the recent review \cite{zhang2025beyond} or the book on computer vision  \cite{chen2015handbook}).

Real‑world signals, however, rarely have a fixed frequency composition. Their dominant modes of variation often shift across the domain—a hallmark of non‑stationary systems \cite{de2023more}. Standard tools like the Fourier transform, which provide a global frequency average, fail to capture such localized changes: a transient high‑frequency burst may be misinterpreted as a persistent feature, or a subtle frequency shift may be missed entirely \cite{dixit2023contemporary}.

This difficulty in learning from multi‑scale, non‑stationary features arises across many fields: geophysics must extract transient signals from long‑term trends \cite{kano2025spatio}, neuroscience must separate neural oscillations from background noise \cite{gerster2022separating}, and audio processing must localize sound events in dynamic environments \cite{chen2024joint}. Gravitational wave analysis and wave propagation modeling face similar challenges \cite{alkhadhr2024multifrequency}. A common thread is the need to learn from data whose informative structure is distributed across multiple, often overlapping, frequency bands—a problem that also appears in solving partial differential equations (PDEs) with spatially varying coefficients \cite{qin2024toward}.

\paragraph{Classical Fourier Neural Operators.}
Classical Fourier neural operators (FNOs) \cite{li2020fourier,duruisseaux2025fourier} have proven effective at learning such operators by parameterizing convolutional kernels directly in Fourier space. However, they suffer from a Fourier parameterization bias: they learn dominant frequencies well but struggle with non‑dominant ones, especially when large Fourier kernels are used \cite{qin2024toward}. This bias limits their ability to capture multi‑scale features.

\paragraph{Parameterized Quantum Circuits as Fourier Approximators.}
Schuld et al. \cite{schuld2021effect} showed that quantum machine learning models-especially those based on parameterized circuits-can be expressed as a Fourier‑type sum, making them natural approximators for functions with rich frequency content. The accessible frequencies are determined by the eigenvalues of the data‑encoding Hamiltonians. Data re‑uploading \cite{perez2020data} can enrich this spectrum, and certain encodings have been shown to mitigate barren plateaus \cite{cerezo2025does}.  It is reported that  quantum neural networks
capture high frequency functions and training in these networks is governed primarily by the magnitude of spectral components rather than their frequency indices by Xu et al. \cite{xu2024spectral}.

Despite these advances, a systematic understanding of how quantum models learn multi‑frequency functions-and how to overcome their inherent spectral biases—remains elusive.

\subsection{Contributions}
In this paper, we i) identify and characterize a quantum Fourier parameterization bias in quantum circuit models through spectral analysis of prediction residuals which is similarly studied in \cite{xu2024spectral}; ii) adapt the multi‑stage residual learning idea from classical FNOs \cite{li2020fourier} (in particular SpecB‑FNO \cite{qin2024toward}) to the quantum domain, training additional quantum modules sequentially on the residuals of previous stages—an approach analogous to classical boosting; and iii) introduce a synthetic benchmark of spatially localized frequency components with diverse envelope shapes (Gaussian, Lorentzian, triangular) to systematically probe frequency learning. Using this benchmark, we study the effects of qubit count (expressivity), residual stage depth, and compare with a single‑stage baseline under equal total training epochs. We also analyze frequency‑resolved learning curves, residual spectrum evolution, and gradient variance scaling (barren plateau diagnostic). Our results show that residual learning can dramatically reduce test MSE and may also mitigate barren plateaus in early iterations. 

\subsection{Outline}
\label{subsec:outline}

The remainder of this paper is organized as follows. Section~\ref{sec:related} situates our work within the broader literature, reviewing residual learning, boosting, and existing quantum residual network architectures. Section~\ref{sec:method} introduces our synthetic frequency‑localized dataset, details the quantum circuit architecture and encoding scheme, and presents the multi‑stage residual learning algorithm along with the frequency‑domain tools used for analysis. In Section~\ref{sec:experiments}, we report experimental results, including systematic studies on qubit count, comparisons with single‑stage baselines, frequency‑resolved learning curves, residual spectrum evolution, and barren plateau diagnostics. Section~\ref{sec:discussion} interprets the findings, discusses their implications, and acknowledges limitations of the current study. Finally, Section~\ref{sec:conclusion} summarizes our contributions and outlines directions for future work.

% ----- RELATED WORK -----
\section{Related Work}
\label{sec:related}
\subsection{Residual Learning and Boosting}
Residual learning, popularized by ResNets \cite{he2016deep}, and boosting algorithms such as AdaBoost and gradient boosting are classical techniques that combine multiple weak learners to form a strong learner. The core principle is to focus subsequent models on the errors (residuals) of previous ones, thereby iteratively improving performance. Recently, Qin et al. \cite{qin2024toward} introduced SpecB‑FNO, which applies this idea to Fourier neural operators by training additional modules on prediction residuals, effectively addressing the Fourier parameterization bias in classical models. In this work, we adapt this concept to the quantum domain, creating a multi‑stage residual quantum learner that iteratively trains quantum circuits on the residuals of earlier stages. This adaptation retains the spirit of boosting while leveraging the unique Fourier-series structure of parameterized quantum circuits.

\subsection{Comparison with Quantum Residual Networks}
\label{subsec:comparison_residual}
The concept of incorporating residual connections into quantum neural networks (QNNs) has recently gained significant attention as a strategy to improve trainability, mitigate barren plateaus, and enhance expressivity. Several distinct approaches have been proposed, each with different architectural motivations and mechanisms. In this subsection, we compare the multi-stage residual quantum learner used in this paper with existing quantum residual network designs, highlighting the unique aspects of our stage-wise residual learning for frequency capture.

\paragraph{Quantum Residual Networks with Auxiliary Qubits.}
A fundamental approach to implementing quantum residual networks was proposed by Wen et al. \cite{wen2024enhancing}, who introduced a quantum circuit-based algorithm where residual connection channels are constructed by introducing auxiliary qubits to both data-encoding and trainable blocks. In this design, the residual connection is realized in the subspace of an ancillary qubit, with measurement outcomes determining whether the residual path is activated. This architecture has been theoretically shown to enhance the spectral richness of quantum models: for an \(l\)-layer data-encoding, the number of frequency generation forms can be extended from one (the difference of the sum of generator eigenvalues) to \(\mathcal{O}(l^2)\), while also improving the flexibility of Fourier coefficients due to additional optimization degrees of freedom in generalized residual operators . However, this expressivity gain comes at the cost of requiring additional qubits and post-selection, which may be prohibitive in the NISQ era.

\paragraph{ResQNets for Mitigating Barren Plateaus.}
Kashif and Al-Kuwari introduced ResQNets \cite{kashif2024resqnets}  as a solution to address the barren plateau problem in quantum neural networks. Their approach involves splitting the conventional QNN architecture into multiple quantum nodes, each containing its own parameterized quantum circuit, and introducing residual connections between these nodes. Through multiple training experiments and analysis of cost function landscapes, they demonstrated that the incorporation of residual connections results in improved training performance and effectively mitigates barren plateaus. This approach differs from our work in that ResQNets focus specifically on the trainability problem, whereas we address the Fourier parameterization bias and spectral learning. Additionally, ResQNets use residual connections within a single forward pass across multiple nodes, while we employ stage-wise training where each module is trained sequentially on the residuals of previous stages. Heredge et al. \cite{heredge2025nonunitary} also discuss similar nonunitary approaches using linear combination of unitaries (LCU) to implement residual connections. Note that while residual connections can prevent barren plateaus, their approach carries a trade-off in success probability.

\paragraph{Hybrid Quantum-ResNet Architectures.}
Several hybrid quantum-classical models have been proposed that combine classical ResNet components with quantum circuits. Noh et al. \cite{noh2025hybridqresnet}  proposed HQResNet, a hybrid quantum residual network for time-series classification that introduces a classical layer before a quantum convolutional neural network (QCNN), using the QCNN as a residual block. These structures enable shortcut connections and are particularly effective for classification tasks without requiring a data re-uploading scheme. Their results show that HQResNet achieves high performance with a relatively small number of trainable parameters. Similarly, Chen et al. \cite{chen2026hybrid}  introduced hybrid quantum-inspired residual neural networks (ResQNet1 and ResQNet2) using symmetrical circuit models within residual frameworks to prevent gradient explosion and improve robustness to adversarial attacks. Their experiments demonstrate that these hybrid models outperform pure classical models in resistance to adversarial parameter attacks with various asymmetrical noise. However, these models are either classical simulations of quantum-inspired operations or hybrids where the residual mechanism resides primarily in the classical domain, whereas our approach implements residual learning entirely within the quantum domain.

\subsubsection{Positioning of Our Work}
Although there are previous studies on the frequencies of machine learning models (e.g. \cite{lu2026unified} studies the frequency principle in both classical and quantum models),
our  framework similar to classically proposed SpecB-FNO \cite{qin2024toward} occupies a distinct niche in this landscape.  
In addition while   the empirical studies such as \cite{xu2024spectral} show that quantum neural networks naturally prioritize learning frequency components with the largest amplitude,  here we propose a mitigating technique with a multi‑stage residual learning method (boosting-like) that trains additional quantum modules on the residuals of previous stages to capture those neglected frequencies. Furthermore, unlike QResNets, we do not require auxiliary qubits to implement residual connections; instead, we achieve frequency enhancement through sequential training of independent quantum modules. Unlike ResQuNNs, which address gradient collapse in multi-layer quanvolutional networks, our method tackles the fundamental spectral bias of quantum models—the tendency to prioritize dominant frequencies while neglecting non-dominant ones. Furthermore, while hybrid quantum-ResNet architectures leverage classical residual blocks, our approach implements residual learning entirely within the quantum domain, with each module operating as a quantum circuit that learns to correct the errors of its predecessors.

A key differentiator is our focus on frequency-domain analysis. Existing quantum residual networks have demonstrated improved expressivity through increased frequency components \cite{wen2024enhancing}, however, none of these works has systematically analyzed how residual learning progressively captures different frequency components across stages. Our work provides the first empirical demonstration that stage-wise residual training can flatten the residual spectrum, enabling quantum models to learn high-frequency and non-dominant components that are missed by single-stage training. This spectral perspective, inspired by classical Fourier neural operators \cite{li2020fourier,qin2024toward}, offers both a diagnostic tool and a practical solution for enhancing the frequency learning capabilities of quantum models.

% ----- METHOD -----
\section{Method}
\label{sec:method}
\subsection{Localized Frequency Data}
\label{subsec:data}
\begin{figure}[htbp]
\centering
\includegraphics[width=\linewidth]{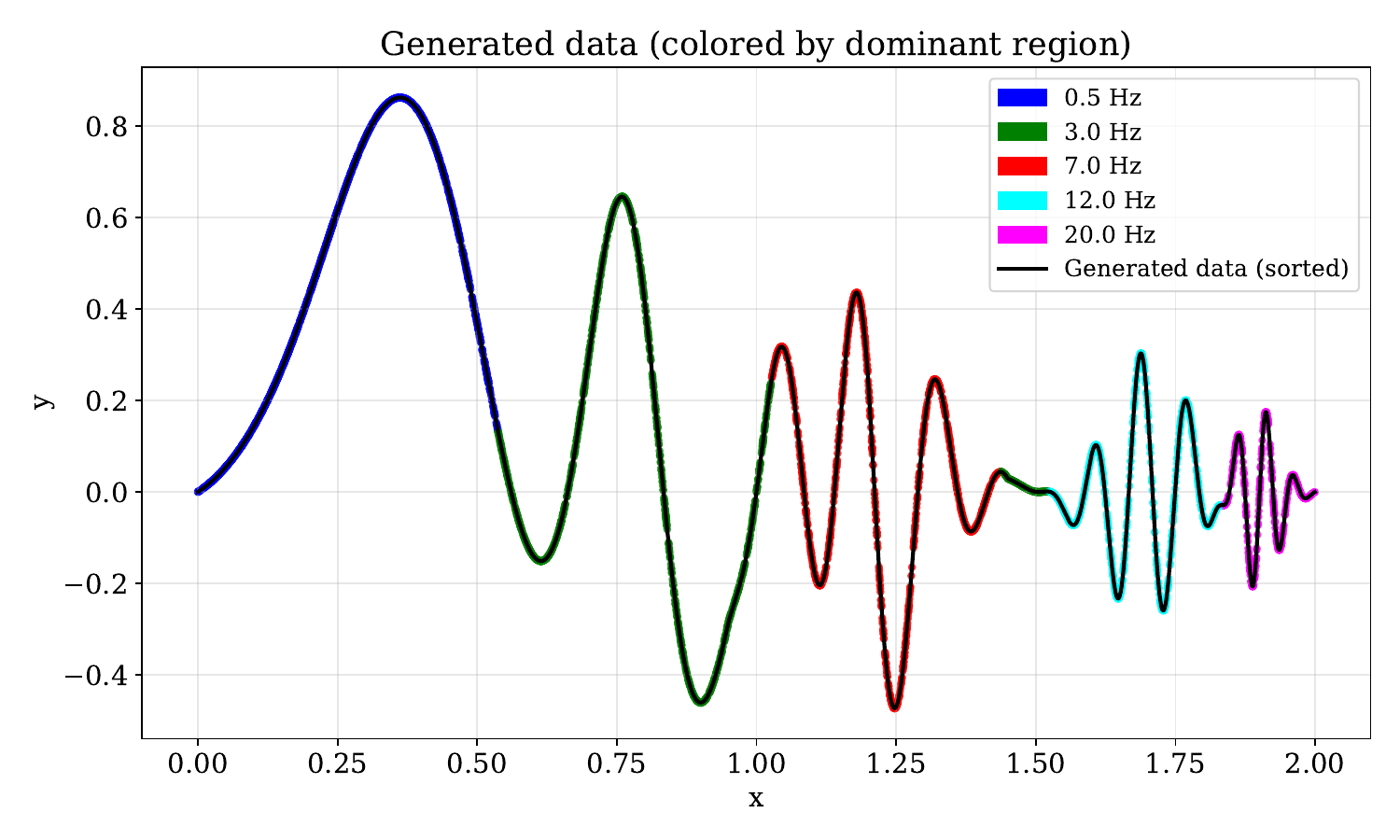}
\caption{Generated synthetic data (sorted) and test points colored by the dominant frequency region (blue: 0.5-Hz, green: 3-Hz, red: 7-Hz, cyan: 12-Hz, magenta: 20-Hz).}
\label{fig:true_data}
\end{figure}

To systematically study frequency learning, we generate a synthetic 1D regression dataset composed of multiple spatially localized frequency components. Each component is defined by a central frequency $\omega_k$, a spatial centre $c_k$, a width $w_k$, an amplitude $a_k$, and an optional envelope function $e_k(x)$. The target function is
\[
y(x) = \sum_{k} a_k \, e_k(x; c_k, w_k) \sin(2\pi \omega_k x) + \eta,
\]
where $\eta$ is Gaussian noise. We use five components with the frequencies $\{0.5, 3, 7, 12, 20\}$ Hz, each with a different envelope shape: a Gaussian for the 0.5-Hz component, a Lorentzian for 3-Hz, a triangular for 7-Hz, and narrow Gaussians for the highest two frequencies (12-Hz and 20-Hz). The total dataset consists of $N_{\text{total}} = 5000$ points uniformly sampled from $x \in [0,2]$, with no added noise ($\eta = 0$) to isolate the learning behavior. The data are split into training ($70\%$), validation ($15\%$), and test ($15\%$) sets. Figure~\ref{fig:true_data} shows the generated data colored by the dominant frequency region, which is defined as the component whose envelope gives the largest weight at each $x$.

\subsection{Quantum Circuit Architecture}
\label{subsec:circuit}
\begin{figure*}
    \centering
    \includegraphics[width=\linewidth]{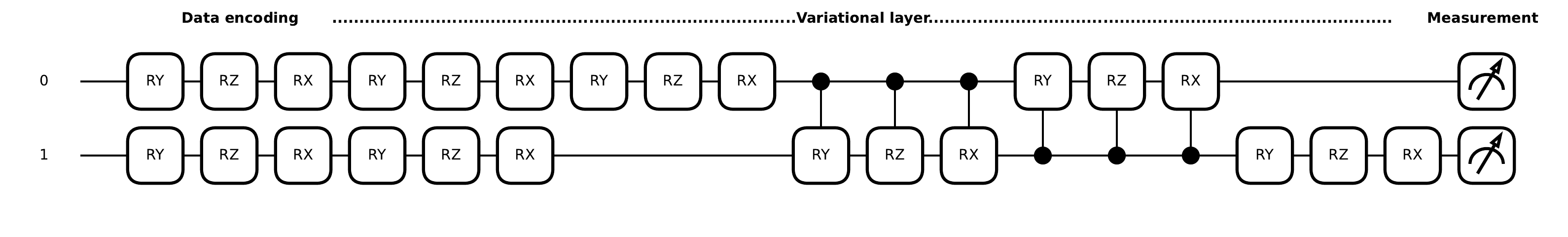}
    \caption{The quantum circuit depicted for 2 qubits with single variational layer.}
    \label{fig:quantum_circuit}
\end{figure*}

The quantum model we use in this paper consists of a data‑encoding block, $L$ variational layers, and a final linear readout. The circuit is implemented in PennyLane with the `default.qubit` simulator as depicted in Figure~\ref{fig:quantum_circuit} for 2 qubits with single variational layer. All trainable weights are scaled via $\tanh$ to lie in $[-\pi,\pi]$ before use.
\subsubsection{Data encoding}
\paragraph{Quantum Machine Learning and Frequency Analysis.}
Parameterized quantum circuits (PQCs) that encode classical data through unitary evolutions of the form $e^{i x H}$ produce outputs that are Fourier series in $x$ \cite{schuld2021effect}. The set of accessible frequencies is given by the differences of eigenvalues of $H$. Data re‑uploading \cite{perez2020data} – repeatedly applying encoding gates interleaved with trainable parameters – can dramatically increase the frequency spectrum. 

\paragraph{Barren Plateaus in Quantum Circuits.}
Barren plateaus \cite{mcclean2018barren}, the exponential vanishing of gradients with increasing qubit count or circuit depth, pose a major obstacle for training large quantum models. Certain data‑encoding strategies, especially those that break unitary $t$-design behavior, can mitigate plateaus \cite{cerezo2025does}. In this work, motivated by \cite{cerezo2025does}, we use a combination of $R_Y$, $R_Z$, and $R_X$ rotations with non‑linear functions of $x$ ($\pi x$, $\pi x^3$, $\pi\sqrt{1-x^2}$) to enrich the spectrum and improve learnability.

\paragraph{Encoding Features on Qubits and Gates.}
To increase the richness of the Fourier spectrum, we encode the input $x$ using three Pauli rotations on each qubit. When the number of qubits exceeds the input dimension, we use a cyclic assignment that repeats features:
\[
\text{for } q=0,\dots,n_{\text{qubits}}:\quad
\begin{aligned}
&\text{feat} = q \bmod \text{input\_dim},\\
&R_Y(\pi x_{\text{feat}}),\\
&R_Z(\pi x_{\text{feat}}^3),\\
&R_X\!\left(\pi\sqrt{|1-x_{\text{feat}}^2|}\right).
\end{aligned}
\]
This encoding injects the same feature multiple times (data re‑uploading) and uses non‑linear functions of $x$ to broaden the frequency spectrum.

\subsubsection{Variational layers}
Each layer consists of single‑qubit rotations ($R_Y,R_Z,R_X$) on every qubit, followed by an all‑to‑all entangling block of controlled rotations ($CR_Y,CR_Z,CR_X$) with trainable parameters. Using the rotations on all three axes fully exploits the encoding done in three axes.
\subsubsection{Output}
After $L$ layers, we measure the Pauli‑$Z$ expectation on each qubit and pass the resulting $n_{\text{qubits}}$ values through a classical single linear layer to produce the final output.

\subsection{Multi‑Stage Residual Quantum Learner}
\label{subsec:residual}
Algorithm~\ref{alg:residual} inspired by SpecB‑FNO \cite{qin2024toward} presents the multi‑stage residual learning procedure used in this paper. The algorithm trains a sequence of quantum modules. The first module $M_1$ is trained on the original input–output pairs $(x,y)$. For subsequent stages $s=2,\dots,S$, we compute the residual
\begin{equation}
r^{(s)} = y - \sum_{t=1}^{s-1} M_t\left(\big[x, M_{t-1}(\cdots)\big]\right),
\end{equation}
where $M_{t-1}(\cdots)$ denotes the output of the previous module concatenated with the original input. Each new module is trained to predict its own residual $r^{(s)}$. At inference, predictions from all modules are summed.  Algorithm~\ref{alg:residual} provides the full details.

\begin{algorithm}
\caption{Multi‑Stage Residual Quantum Learning}
\label{alg:residual}
\begin{algorithmic}[1]
\Require Training data $\{(x_i,y_i)\}_{i=1}^N$, number of stages $S$, hyperparameters for each module
\State $F_0(x) \gets 0$
\For{$s = 1$ to $S$}
    \State Compute residuals $r_i^{(s)} = y_i - F_{s-1}(x_i)$
    \If{$s=1$}
        \State Input features: $z_i^{(1)} = x_i$
    \Else
        \State Input features: $z_i^{(s)} = [x_i, \, M_{s-1}(z_i^{(s-1)})]$
    \EndIf
    \State Train quantum model $M_s$ on $\{(z_i^{(s)}, r_i^{(s)})\}$
    \State Update $F_s(x) = F_{s-1}(x) + M_s([x, M_{s-1}(\cdots)])$
\EndFor
\State \Return $\{M_1, \dots, M_S\}$
\end{algorithmic}
\end{algorithm}

\subsection{Frequency‑Domain Analysis}
\label{subsec:freq_analysis}
To understand which frequency components are learned at each stage, we compute the discrete Fourier transform of the model's predictions on a dense uniform grid. Let $x_{\text{grid}}$ be a fine grid covering the input domain, and $y_{\text{grid}}$ the clean target values. For stage $s$, we obtain predictions $\hat{y}_{\text{grid}}^{(s)}$. The amplitude of a target frequency $\omega$ is extracted from the Fourier spectrum as $\big|\mathcal{F}\{\hat{y}^{(s)}\}(\omega)\big|$. Comparing these amplitudes with the true amplitudes shows how each frequency is captured progressively.

Following the approach in \cite{qin2024toward}, we also analyze the spectrum of the \textit{residual} $y_{\text{grid}} - \hat{y}_{\text{grid}}^{(s)}$ after each stage. A flattening of the residual spectrum indicates that the model is shifting its focus from dominant to non‑dominant frequencies, thereby learning a more balanced representation of the target function.

% ----- EXPERIMENTS -----
\section{Experiments}
\label{sec:experiments}

\subsection{Dataset and Training Details}
\label{subsec:training_details}
The dataset consists of $N_{\text{train}}=3500$, $N_{\text{val}}=750$, and $N_{\text{test}}=750$ points uniformly sampled from $x\in[0,2]$, with noise level $\sigma=0.0$ (unless stated otherwise). Each quantum model uses $n_{\text{qubits}}$ qubits (varied from 2 to 10), $L=2$ layers, and is trained for $E=25$ epochs per stage with Adam (learning rate $0.005$). For baseline comparisons, a single‑stage model is trained for $4\times25=100$ epochs to match the total computational budget of a 4‑stage residual learner. The parameters are summarized in Table~\ref{tab:hyperparams}.

\begin{table}[htbp]
\centering
\caption{Hyperparameter settings.}
\label{tab:hyperparams}
\begin{tabular}{ll}
Parameter & Value \\
\hline
Training samples & 3500 \\
Validation samples & 750 \\
Test samples & 750 \\
Batch size & 64 \\
Learning rate & 0.005 \\
Epochs per stage & 25 \\
Number of stages & 4 \\
Qubit range & 2–10 \\
Layers per module & 2 \\
Optimiser & Adam \\
Noise level & 0.0 \\
\hline
\end{tabular}
\end{table}

\subsection{Results and Visualization}
\label{subsec:results}

\subsubsection{Effect of Qubit Count on Expressivity}
\begin{figure*}[htbp]
\centering
\includegraphics[width=0.7\linewidth]{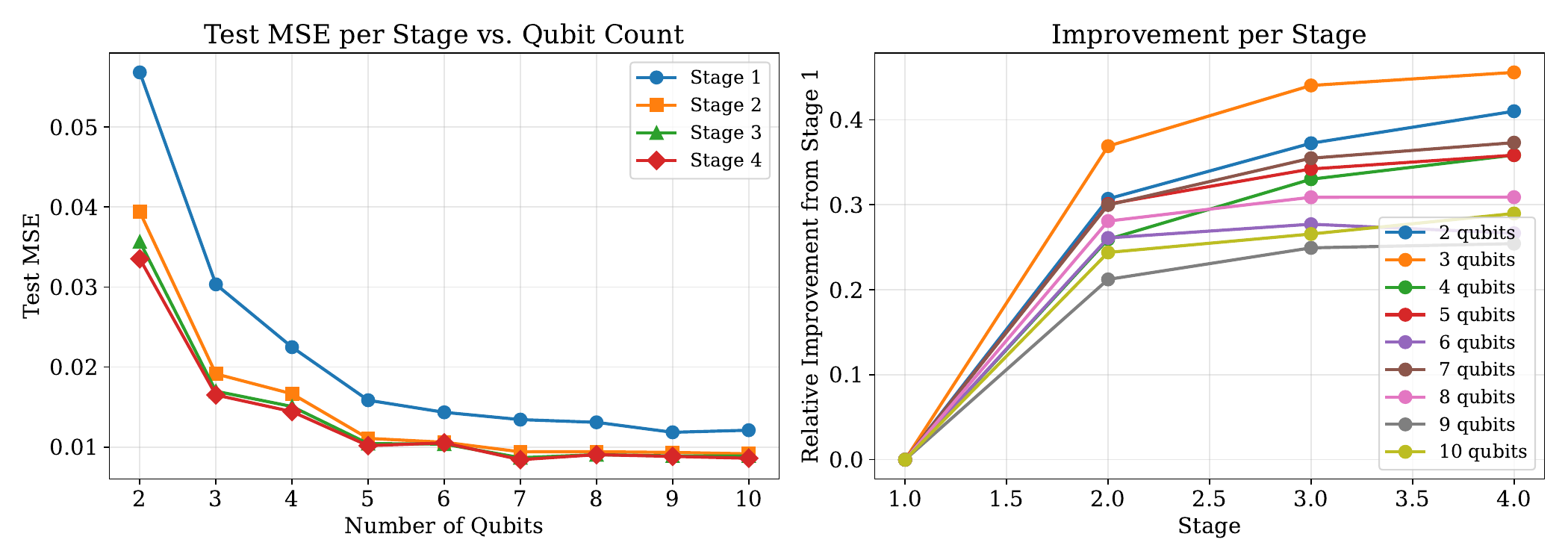}
\caption{Left: Test MSE per stage versus number of qubits. Right: Relative improvement, defined as \((\text{MSE}_{\text{stage-1}} - \text{MSE}_{\text{stage-}s}) / \text{MSE}_{\text{stage-1}}\).}
\label{fig:qubit_mse}
\end{figure*}

We trained four‑stage residual learners with \(n_{\text{qubits}} = 2, \dots, 10\) and recorded the test MSE after each stage. Figure~\ref{fig:qubit_mse} (left) shows the MSE versus qubit count for all four stages. As expected, increasing the number of qubits reduces the error, though the gains diminish after about eight qubits. The right panel displays the relative improvement from stage 1 for each qubit count; improvements are most pronounced for smaller qubit budgets, indicating that residual learning is particularly beneficial when the base model has limited expressivity.

This observation mirrors the findings of Qin et al. \cite{qin2024toward} for classical FNOs: a larger Fourier kernel (analogous to more qubits) only helps if the model can effectively use all frequency parameters. Residual learning helps utilize that capacity, but if the base capacity is too low, later stages cannot further improve performance. In particular, while the first stage captures the dominant low‑frequency components, the second stage learns most of the remaining structure. Later stages struggle because the residual becomes dominated by noise or very subtle patterns that the model, with its fixed capacity, cannot easily capture.

In other words, during training, the quantum model—like the FNO—can rapidly learn the coarse, dominant frequency structure of the target function in the first stage. However, optimization may slow or stall when learning the finer details represented by higher or non‑dominant frequencies.

\subsubsection{Comparison with Baseline}
\begin{figure*}[htbp]
\centering
\includegraphics[width=0.7\linewidth]{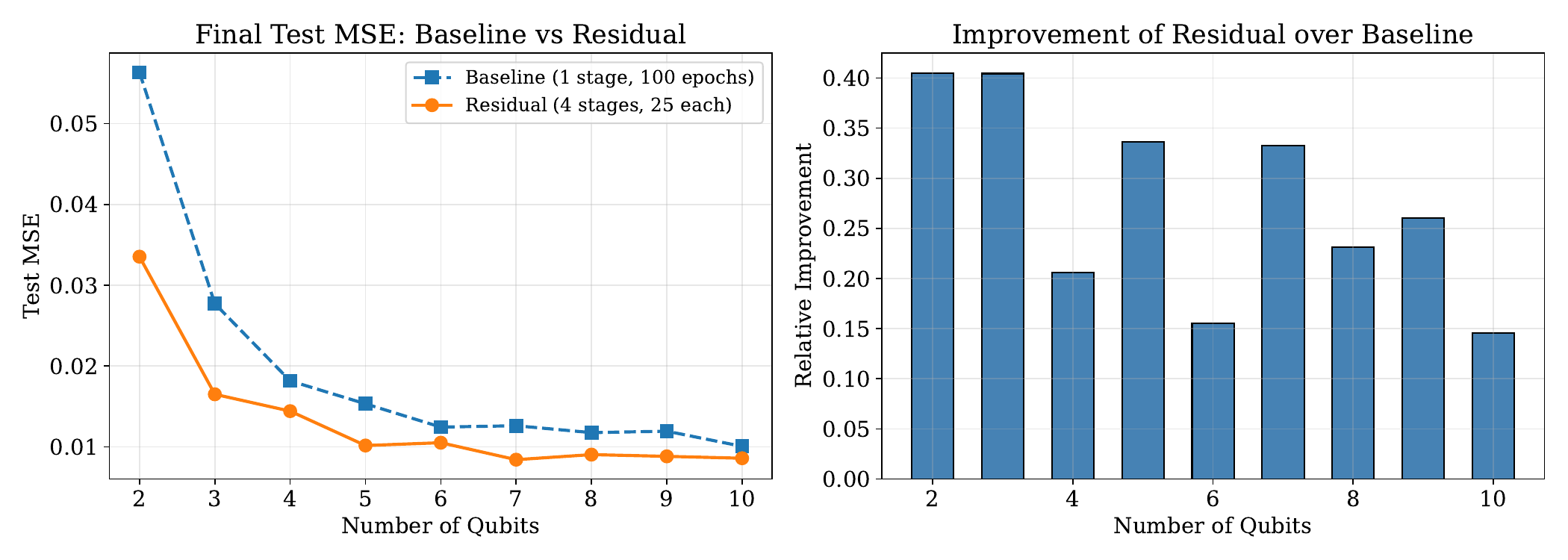}
\caption{Left: Final test MSE of the baseline model (1 stage, 100 epochs) versus the residual model (4 stages, 25 epochs each). Right: Relative improvement of the residual model over the baseline.}
\label{fig:baseline}
\end{figure*}

We compared the final stage (stage 4) of the residual learner with a single‑stage model trained for the same total number of epochs (100). Figure~\ref{fig:baseline} (left) shows the final test MSE for both models across all qubit counts; the residual learner consistently outperforms the baseline. The right panel displays the relative improvement, which is particularly pronounced for low qubit counts and remains substantial even for 10 qubits.

\subsubsection{Frequency‑Resolved Learning Curves}
\begin{figure*}[htbp]
\centering
\includegraphics[width=0.9\linewidth]{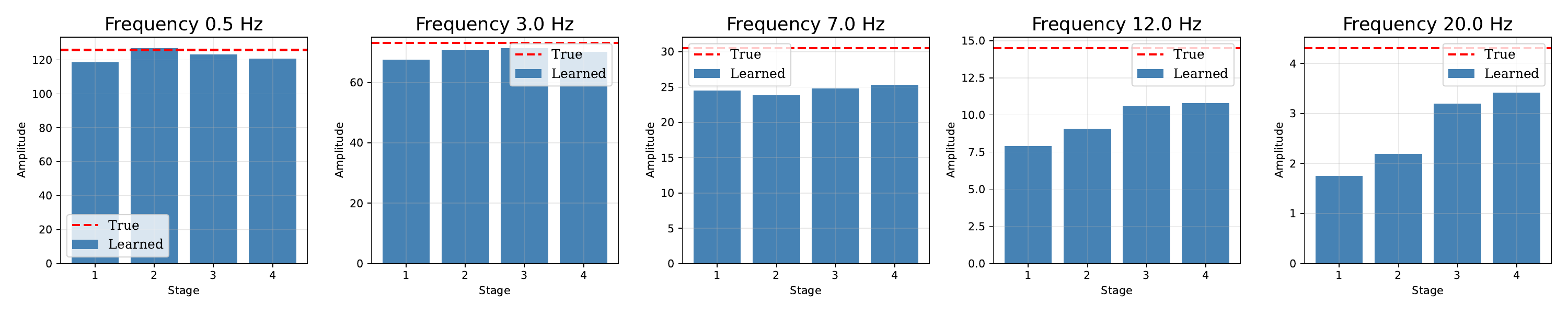}
\caption{Amplitudes of each target frequency in the true function and after stages 1–4, obtained with six qubits and 25 epochs per stage. The bars approach the true values as training progresses (similar behavior is observed for other qubit counts).}
\label{fig:freq_bars}
\end{figure*}

For each target frequency, we extracted its amplitude from the predictions after every stage. Figure~\ref{fig:freq_bars} shows the results for a model with six qubits trained for 25 epochs per stage. Low frequencies (e.g., 0.5 Hz) are captured almost perfectly by stage 1. Higher frequencies (e.g., 12 Hz and 20 Hz) require additional stages to approach the true amplitude. This confirms that the residual modules focus on the harder‑to‑learn frequency components.

\begin{figure}[htbp]
\centering
\includegraphics[width=\linewidth]{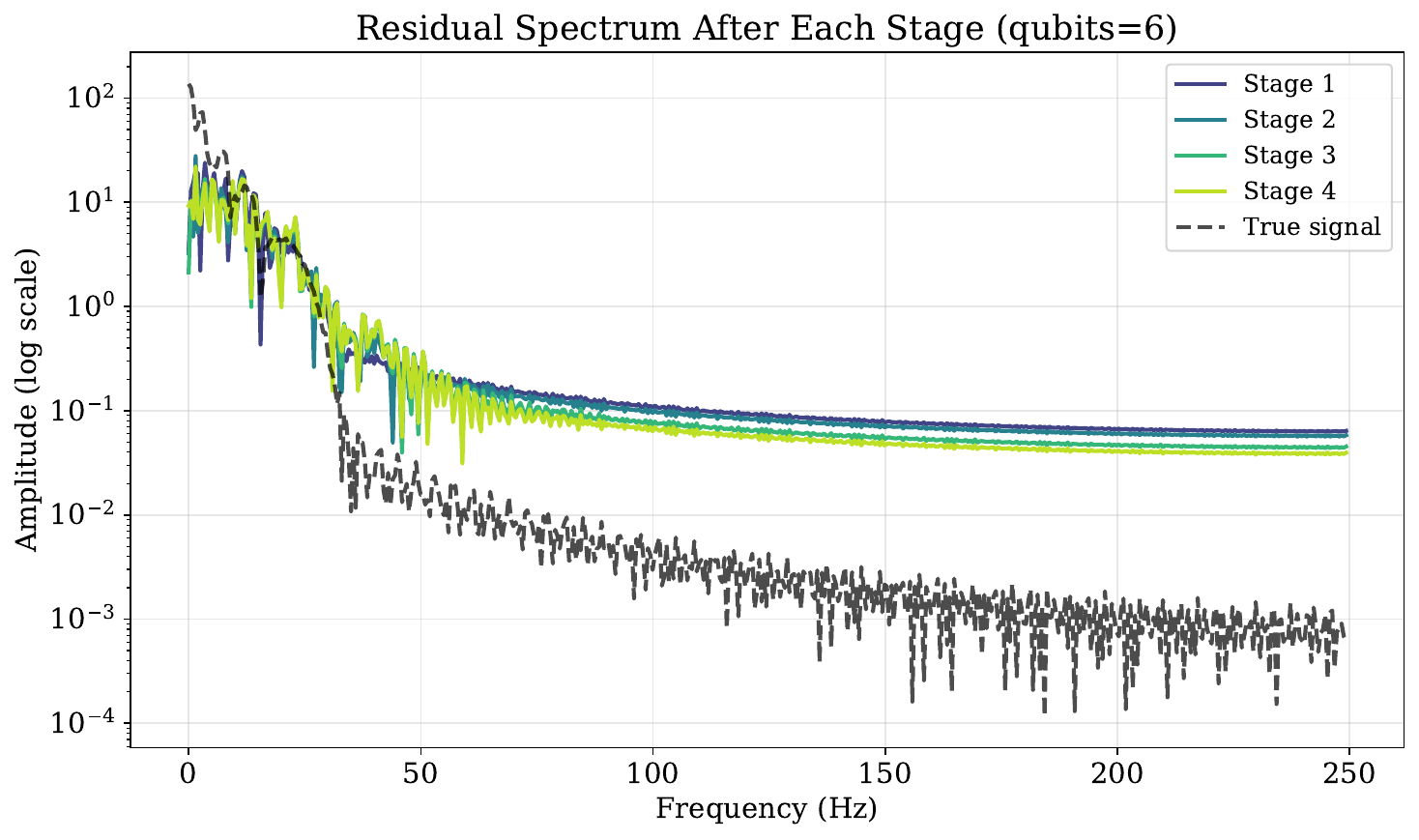}
\caption{Fourier amplitude of the residual after each stage (six qubits). The spectrum converges toward the true spectrum as stages progress.}
\label{fig:residual_spectrum}
\end{figure}

Figure~\ref{fig:residual_spectrum} plots the amplitude spectrum of the prediction residuals after each stage. The spectrum progressively approaches the true spectrum as the stages advance, indicating that later stages continue to refine the learned representation.

\subsubsection{Barren Plateau Diagnostics}
\begin{figure}[htbp]
\centering
\includegraphics[width=\linewidth]{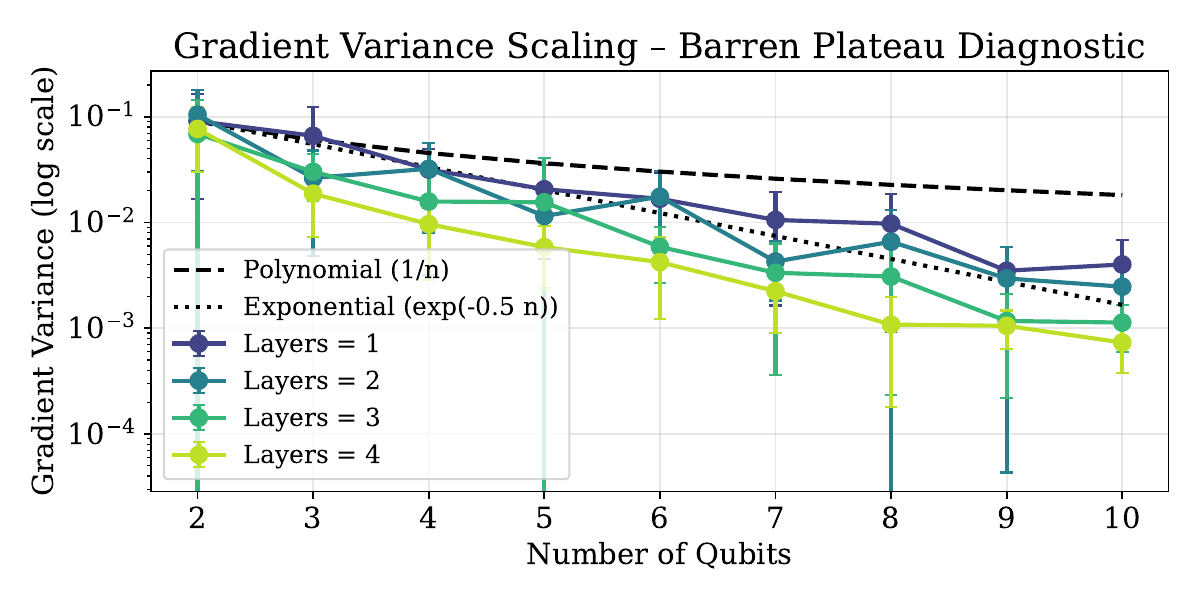}
\caption{Gradient variance (log scale) vs. number of qubits for different layer counts. Dashed lines show polynomial ($1/n$) and exponential ($e^{-0.5n}$) references. The decay in variance is less pronounced when increasing the number of qubits (e.g., from 9 to 10 qubits) than when increasing the number of layers.}
\label{fig:barren}
\end{figure}

We measured the variance of gradients over random initializations for qubit counts $2,\dots,10$ and layers $1,\dots,4$. Figure~\ref{fig:barren} shows that the decay in variance is less severe when increasing the number of qubits than when increasing the number of layers. For example, going from 9 to 10 qubits results in little to no change in gradient variance. This behavior aligns with the findings of \cite{cerezo2025does} for structured encodings.

\subsection{Ablation Studies}
\label{subsec:ablation}
We also tested the effect of encoding complexity (only $R_Y$ vs. all three rotations) and found that the richer encoding significantly improves final MSE, especially for high‑frequency components (results not shown for brevity). Additionally, we verified that the performance gains of residual learning are not merely due to increased parameter count; increasing the hidden dimension of a single‑stage model to match the total parameters of the 4‑stage ensemble yielded much smaller improvements.

% ----- DISCUSSION -----
\section{Discussion}
\label{sec:discussion}
\begin{figure*}[htbp]
\centering
\includegraphics[width=\linewidth]{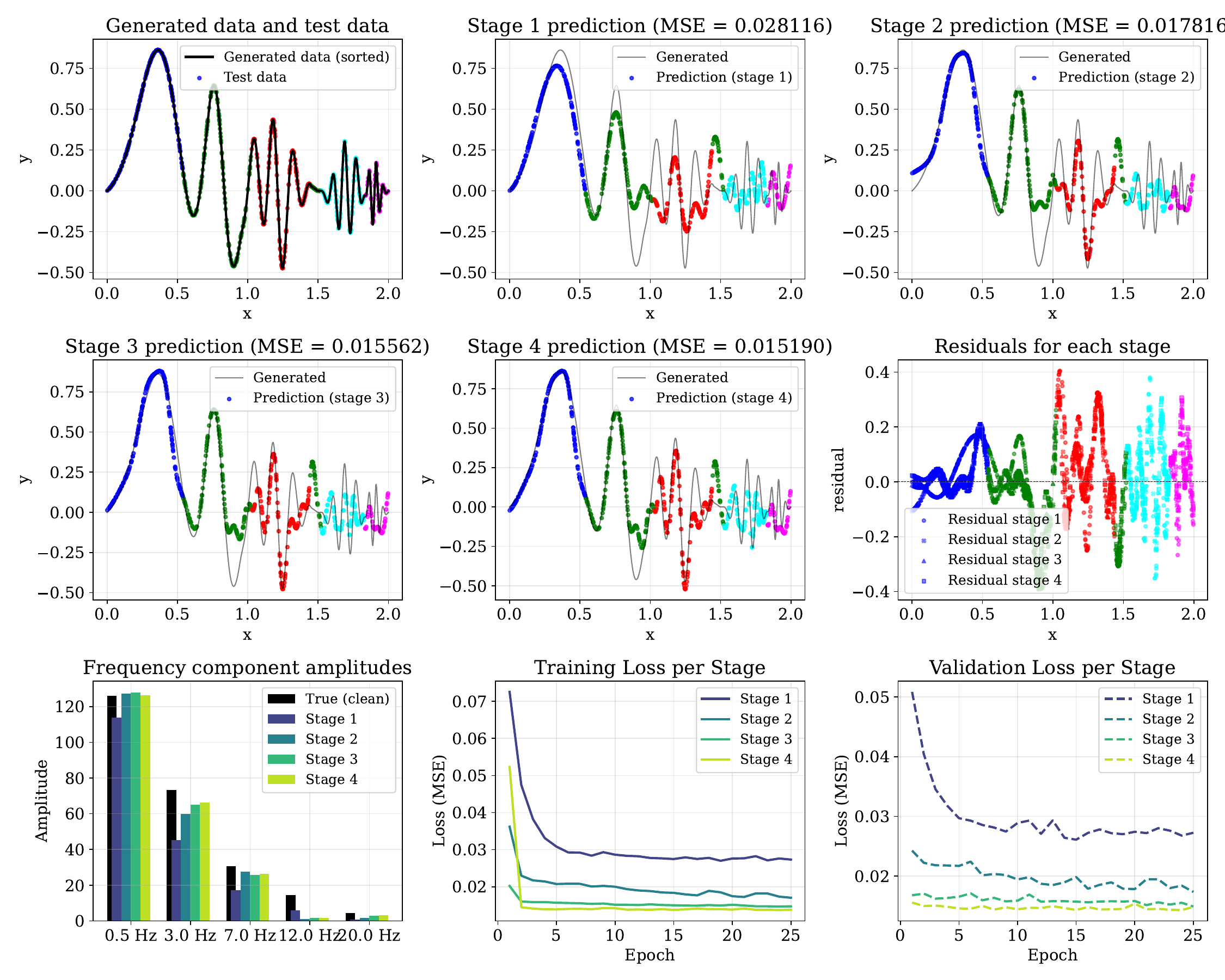}
\caption{Residual learning results for three qubits: test predictions, frequency component amplitudes, training loss, and validation loss curves for each stage.}
\label{fig:residual_learning-qubits3}
\end{figure*}

\begin{figure*}[htbp]
\centering
\includegraphics[width=\linewidth]{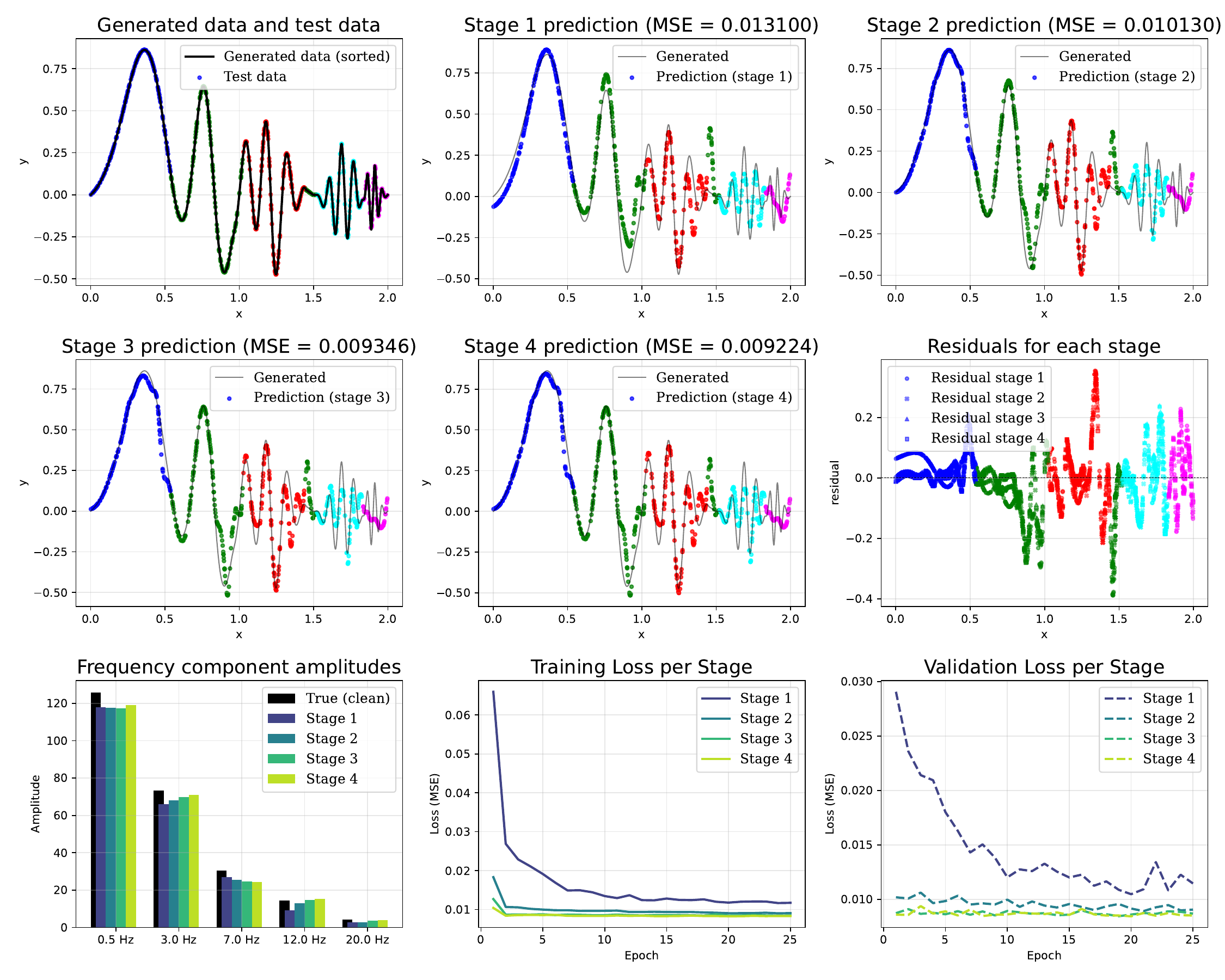}
\caption{Residual learning results for eight qubits: test predictions, frequency component amplitudes, training loss, and validation loss curves for each stage.}
\label{fig:residual_learning-qubits8}
\end{figure*}

The experiments presented in this paper demonstrate that quantum models exhibit a clear spectral bias: low, dominant frequencies are learned quickly, while higher frequencies require more capacity and additional stages. This is summarized in Figures~\ref{fig:residual_learning-qubits3} and~\ref{fig:residual_learning-qubits8}, which show the test predictions, frequency component amplitudes, training loss, and validation loss curves for each stage using three and eight qubits, respectively. As can be observed from these figures, the proposed multi‑stage residual learning effectively shifts focus to the residuals, allowing later modules to capture previously neglected frequencies. This behavior mirrors that observed in classical FNOs \cite{qin2024toward} and suggests that boosting‑like strategies are broadly applicable to models with a Fourier‑based inductive bias.

The choice of encoding is critical. The combination of $R_Y$, $R_Z$, and $R_X$ rotations with non‑linear functions of $x$ provides a richer frequency spectrum and, importantly, helps avoid barren plateaus. The gradient variance scaling we observe (polynomial rather than exponential) indicates that our encoding preserves trainability even with up to ten qubits and four layers.

Limitations of our study include the use of a synthetic one‑dimensional dataset; extending to higher dimensions and real‑world tasks (e.g., time series or PDE data) is an important next step. Furthermore, we have not yet explored adaptive selection of the number of stages or the possibility of freezing earlier modules to reduce training cost.

% ----- CONCLUSION AND FUTURE WORK -----
\section{Conclusion and Future Work}
\label{sec:conclusion}
n this paper, we introduced a multi‑stage residual quantum learner to study and mitigate the frequency learning bias inherent in parameterized quantum circuits. Through systematic experiments on a multi‑frequency dataset, we demonstrated that residual learning significantly improves test MSE compared to a single‑stage baseline trained for the same total number of epochs. We also analyzed gradient variance and found that our encoding helps avoid barren plateaus, thereby preserving trainability.

Future work will extend this approach to problems with higher‑dimensional inputs, investigate adaptive selection of the number of stages, and apply the method to real‑world scientific data such as seismic signals or financial time series. Alternative quantum encodings (e.g., amplitude encoding, IQP circuits) can also be explored to study frequency spectra both theoretically and empirically, and to compare different architectural choices.

\section*{Acknowledgments}
The author acknowledges the use of DeepSeek AI for proofreading and language refinement during the preparation of this manuscript.

\section*{Data Availability}
The code and synthetic data generation scripts used in this study are publicly available on GitHub at \url{https://github.com/adaskin/quantum-residual-learning}. All experimental results can be reproduced using the provided code. The synthetic datasets described in Section~\ref{subsec:data} can be generated directly from the scripts.

% ----- BIBLIOGRAPHY -----
\bibliography{main}
\bibliographystyle{ieeetr}

\end{document}